# Theoretical study of the effective modulus of a composite considering the orientation distribution of the fillers and the interfacial damage

Sangryun Lee[1] and Seunghwa Ryu[1,*]

**Affiliations**

[1] Department of Mechanical Engineering, Korea Advanced Institute of Science and Technology (KAIST), 291 Daehak-ro, Yuseong-gu, Daejeon 34141, Republic of Korea

[*] Corresponding author e-mail: ryush@kaist.ac.kr

**Abstract**

In the manufacturing process of a filler-reinforced composite, the fillers are partially aligned due to the shear flow in the drawing stage. Besides, various imperfections form at the interface between the matrix and the fillers, leading to debonding and slip under mechanical loading. There have been numerous micromechanics studies to predict effective moduli of the composites in the presence of partial alignment of fillers and interface imperfections. Here, we present an improved theory that overcomes two limitations in the existing micromechanics based approaches. First, we find that the interface damage tensor for axisymmetric ellipsoidal inhomogeneity developed to model the interfacial damage leads to the prediction of infinite or negative effective moduli. We show that these anomalies can be eliminated if correctly derived damage tensor is used. Second, we reveal that the previous theory on the effective moduli with axisymmetric filler orientation distribution fails because longitudinal and transverse moduli predictions do not converge in the limit of random orientation distribution. With appropriate corrections, we derive analytic expressions for the orientation average of arbitrary transversely isotropic $4^{th}$ order tensor under general axisymmetric orientation distribution. We apply the improved method to compute the effective moduli of a polymer-carbon nanotube composite with non-uniform filler orientation and interface damage.

## 1. Introduction

Due to remarkable mechanical, electrical, and thermal properties, nanocomposites reinforced with carbon nanotubes (CNTs), graphene, and nanowires have attracted considerable attention because of their numerous potential applications, including flexible sensors (Amjadi et al., 2014; Lee et al., 2015), lithium batteries (Cui et al., 2008; Huang et al., 2002; Oh et al., 2010), optical device (Lü et al., 2006; Lü et al., 2005), and flexible energy storage devices (Nyholm et al., 2011; Pushparaj et al., 2007). To effectively design and make use of the nanocomposites, it is essential to accurately predict the effective properties of the composite as a function of shape, volume fraction, and orientation distribution of fillers.

For example, the piezoresistivity of the metal nanowire-polydimenthylsiloxane (PDMS) composites has been utilized for the highly stretchable strain sensors (Amjadi et al., 2014). The effective electrical property of such composites has been analyzed by solving three dimensional electric circuit constructed from a percolation network of thousands of nanowires (Lee et al., 2015). For other examples, the finite element method (FEM) was used to compute the effective moduli of composites at the bulk scale (Banerjee and Sankar, 2014; Sodhani and Reese, 2014; Sun et al., 2001), whereas molecular dynamics (MD) simulations were used at the atomic scale (Alian et al., 2015; Yang et al., 2013a; Yang et al., 2013b). The above examples require computationally expensive and time-consuming calculations to consider large simulation cells involving many fillers to serve as representative volume elements (RVEs).

To compute the effective properties more efficiently, various micromechanics-based approaches have been used, such as the Eshelby method, the Mori-Tanaka (MT) method, and the Self-Consistent (SC) method. Eshelby was the first to introduce the Eshelby tensor ($S$) and to solve the single inhomogeneity in the matrix by converting the problem into an equivalent eigenstrain problem (Eshelby, 1957). The $S$ tensor, which is a 4$^{th}$ order tensor that depends on the matrix properties and the shape of fillers, relates the eigenstrain of inclusion ($\varepsilon^*$) and the constrained strain of the inclusion ($\varepsilon_0$) ($\varepsilon_0 = S:\varepsilon^*$) at single inclusion problem. As extensions of the approach, the MT method and the SC method were devised; these methods are mean-field homogenization schemes to treat the multiple inhomogeneity problems (Hill, 1965; Qiu and Weng, 1990; Withers et al., 1989). The MT method is the most popular method because it provides more accurate predictions on the effective properties than the Eshelby method and has an explicit closed-form solution, whereas the SC method relies on implicit equations. However, the original MT method has two limitations: first, it does not account for the imperfections in the filler-matrix interface, such as debonding and slip; second, it is only applicable when all fillers in the matrix are aligned perfectly along one direction.

To take into account the reduced effective modulus from the interfacial damage, Qu et al. introduced the interface spring model, which assumes a virtual linear spring at the interface to allow for a displacement jump across the interface (Qu, 1993). The 4[th] order interfacial damage tensor ($R$ tensor) (which is referred to as $H$ tensor in Qu et al.(Qu, 1993)) was defined to represent the degree of damage as a function of the filler shape and the compliance of the interface springs. The modified Eshelby tensor is then obtained by using the $R$ tensor, which eventually leads to the change in the effective modulus prediction. Barai et al. firstly derived the closed form of the $R$ tensor for the prolate shape filler (Barai and Weng, 2011). Because the shape of most fillers (such as spherical nanoparticles or carbon nanotube) can be approximated as axisymmetric ellipsoids with varying aspect ratio, the approach has been used to predict the effective moduli of various composites (Pan et al., 2013; Shokrieh et al., 2016; Yang et al., 2013b). Although incorrectly derived $R$ tensor leads to the infinite or negative effective moduli for some ranges of filler aspect ratio, those anomalies have never been recognized in the previous studies (Pan et al., 2013; Shokrieh et al., 2016; Yang et al., 2013b).

The fillers in the matrix are partially aligned because of the shear flow in the drawing stage (Fan and Advani, 2005; Fu et al., 2000; Pötschke et al., 2005). Hence, it is crucial to predict the effective modulus as a function of the filler orientation distribution (Buck et al., 2015). Odegard et al. suggested the orientation average of $4^{th}$ order stiffness tensor using the $3-1-3$ Euler angle and showed that the effective moduli become anisotropic when the fillers are partially or fully aligned (Odegard et al., 2003). However, we find that their prediction on the effective moduli with axisymmetric filler orientation distribution fails because the longitudinal and transverse moduli do not converge in the limit of random orientation distribution due to the geometrical constraint of the Euler angle scheme. Meantime, the orientation average scheme of Odegard et al. has been applied to the study of the effective moduli of composites with various types (Alian et al., 2015; Nguyen et al., 2013; Pan et al., 2016).

To date, researchers have applied the micromechanics to predict the effective moduli of composites for either random or fully aligned orientation distribution of fillers when accounting for the interfacial damage (Dinzart and Sabar, 2017; Kundalwal and Kumar, 2016; Pan et al., 2013). For the studies considering realistic partial alignment of fillers (Dunn et al., 1996; Jiang et al., 2007; Nguyen et al., 2013), the imperfect interfacial bonding has not been considered. In this work, we propose an improved micromechanics model by correcting the two problems regarding the interface damage and the orientation average, and provide an efficient way of computing the effective moduli of composites for general axisymmetric orientation distribution of fillers. We demonstrate that the singularities in effective modulus prediction can be removed when the corrected $R$ tensor is used. We obtain the closed form of $R$ tensor for both prolate and oblate shape fillers and validate our results against the numerical integration results. We also confirm that our expressions of $R$ tensor satisfy two limiting cases, i.e., aspect ratio of 1 and infinity, at which analytic forms are readily available (Qu, 1993). Instead of the $3-1-3$ Euler angle, we use polar and azimuthal angles on the unit sphere to represent the filler orientation and derive algebraic expressions of orientation average for the general transversely isotropic 4$^{\text{th}}$ order tensor under axisymmetric filler orientation distribution. We confirm that the longitudinal and transverse elastic moduli obtained from the correct $R$ tensor and orientation average scheme converge in the random orientation distribution limit. Our results can be widely used to describe composites that include particles and fillers at various aspect ratios.

## 2. Micromechanics

### 2.1 Interfacial damage modeling

We adopt Hill's notation to calculate the effective modulus of RVE which is depicted in Fig. 1. Walpole's scheme is used to perform the complex calculations of 4$^{th}$ order tensors, such as double inner product and inverse conveniently (Qiu and Weng, 1990). In what follows, the capital sized alphabets represent 4$^{th}$ order tensors, and the colons denote double inner product. The effective stiffness tensor of the composite in the original MT approach is expressed as below.

$$L_{eff} = (c_0 L_0 + c_1 L_1 : A) : (c_0 I + c_1 A)^{-1} \tag{1}$$

where $L_0$ and $L_1$ are the stiffness tensors of the matrix and filler, respectively. $I$ is the symmetric identity tensor. $c_0$ and $c_1$ refer to the volume fraction of matrix and fillers, respectively; thus, $c_0 + c_1 = 1$. In MT method, the tensor $A$ is the local strain concentration tensor which relates the volume averaged strains in matrix ($\bar{\varepsilon}_0$) and fillers ($\bar{\varepsilon}_1$) by the definition of $\bar{\varepsilon}_1 \equiv A : \bar{\varepsilon}_0$. The local strain concentration tensor is obtained from a single inhomogeneity problem by comparing strain within the inhomogeneity and external strain ($\varepsilon_1 = A : \varepsilon_{ext}$). We do not need volume average here because the strain within the inhomogeneity is uniform. After solving linear elasticity for the single inhomogeneity problem, the $A$ tensor can be expressed in terms of Eshelby tensor ($S$) and stiffness tensor of each phases ($L_0, L_1$), as Eq.(2).

$$A = [I + S : L_0^{-1} : (L_1 - L_0)]^{-1} \tag{2}$$

The Eshelby tensor ($S$) for the prolate and oblate inclusions has been derived in the literature (Qiu and Weng, 1990), as summarized in the Appendix.

Because the original MT method is only applicable when the fillers are completely aligned and have perfect bonding with the matrix, the modified Mori-Tanaka (mMT) approach must be employed to account for the imperfect bonding at the interface (slip or debonding). To model the interfacial damage, we consider the displacement jump across the interface by adopting the linear spring model (Qu, 1993) (see Fig.2),

$$\Delta u_i = \eta_{ij}\sigma_{jk}n_k, \qquad \eta_{ij} = \alpha\delta_{ij} + (\beta - \alpha)n_i n_j \qquad (3)$$

where the $\eta_{ij}$ tensor refers to the compliance of the interface spring in the tangential ($\alpha$) and normal ($\beta$) directions. The $n_i$ represents outward direction unit normal vector at the inclusion surface. After solving single inclusion problem with the displacement jump, the modified Eshelby tensor($\tilde{S}$) is given as follows (Qu and Cherkaoui, 2006),

$$\tilde{S} = S + S:R:L_0:(I - S). \qquad (4)$$

In this work, we limit our focus on the axisymmetric (prolate and oblate) ellipsoidal fillers with two semi-axial lengths of $a_1$ along the $x_1$ direction and $a$ along the $x_2$ and $x_3$ directions. Due to the nonphysical overlapping arising from a nonzero $\beta$ (Luding, 2008), the $R$ tensor is typically expressed by considering only tangential spring compliance as

$$R_{ijkl} = \alpha(P_{ijkl} - Q_{ijkl}) \qquad (5)$$

where

$$P_{ijkl} = \frac{3}{16\pi}\int_0^\pi\left[\int_0^{2\pi}(\delta_{ik}\widehat{n_j n_k} + \delta_{jk}\widehat{n_i n_l} + \delta_{il}\widehat{n_k n_j} + \delta_{jl}\widehat{n_k n_i})n^{-1}d\chi\right]sin\psi d\psi \qquad (6)$$

and

$$Q_{ijkl} = \frac{3}{4\pi}\int_0^\pi\left[\int_0^{2\pi}\widehat{n_i}\widehat{n_j}\widehat{n_k}\widehat{n_l}n^{-3}d\chi\right]sin\psi d\psi. \qquad (7)$$

with $\widehat{\boldsymbol{n}} = \left(\frac{cos\psi}{a_1}, \frac{sin\psi cos\chi}{a}, \frac{sin\psi sin\chi}{a}\right)^T, n = \sqrt{\widehat{n_i}\widehat{n_i}} = \frac{1}{\rho a}\sqrt{cos^2\psi + \rho^2 sin^2\psi}$, and $\rho = \frac{a_1}{a}$. $\psi$ and $\chi$ are two parameters that represent the surface integral domain, i.e. the surface of the ellipsoidal inclusion. The non-zero independent components of $P$ and $Q$ can be obtained from the integral given below:

$$P_{1111} = \frac{3}{2a}\left[\frac{\rho}{(\rho^2-1)^{3/2}}\sin^{-1}\frac{\sqrt{\rho^2-1}}{\rho} - \frac{1}{\rho(\rho^2-1)}\right],$$

$$P_{2222} = \frac{3}{4a}\left[\frac{\rho(\rho^2-2)}{(\rho^2-1)^{3/2}}\sin^{-1}\frac{\sqrt{\rho^2-1}}{\rho} + \frac{\rho}{(\rho^2-1)}\right]$$

$$Q_{1111} = \frac{3}{2a}\left[\frac{2\rho^2+1}{\rho(\rho^2-1)^2} - \frac{3\rho}{(\rho^2-1)^{5/2}}\sin^{-1}\frac{\sqrt{\rho^2-1}}{\rho}\right]$$

$$Q_{1122} = \frac{3}{4a}\left[\frac{\rho(\rho^2+2)}{(\rho^2-1)^{5/2}}\sin^{-1}\frac{\sqrt{\rho^2-1}}{\rho} - \frac{3\rho}{(\rho^2-1)^2}\right]$$

$$Q_{2222} = \frac{9}{16a}\left[\frac{\rho(2+\rho^2)}{(\rho^2-1)^2} + \frac{\rho^3(\rho^2-4)}{(\rho^2-1)^{5/2}}\sin^{-1}\frac{\sqrt{\rho^2-1}}{\rho}\right]$$

for prolate shape fillers($\rho > 1$), and

$$P_{1111} = \frac{3}{2a}\left[-\frac{\rho}{(1-\rho^2)^{3/2}}\sinh^{-1}\frac{\sqrt{1-\rho^2}}{\rho} + \frac{1}{\rho(1-\rho^2)}\right], \qquad (8)$$

$$P_{2222} = \frac{3}{4a}\left[\frac{\rho(2-\rho^2)}{(1-\rho^2)^{3/2}}\sinh^{-1}\frac{\sqrt{1-\rho^2}}{\rho} - \frac{\rho}{(1-\rho^2)}\right]$$

$$Q_{1111} = \frac{3}{2a}\left[\frac{2\rho^2+1}{\rho(1-\rho^2)^2} - \frac{3\rho}{(1-\rho^2)^{5/2}}\sinh^{-1}\frac{\sqrt{1-\rho^2}}{\rho}\right]$$

$$Q_{1122} = \frac{3}{4a}\left[\frac{\rho(\rho^2+2)}{(1-\rho^2)^{5/2}}\sinh^{-1}\frac{\sqrt{1-\rho^2}}{\rho} - \frac{3\rho}{(1-\rho^2)^2}\right]$$

$$Q_{2222} = \frac{9}{16a}\left[\frac{\rho(2+\rho^2)}{(1-\rho^2)^2} - \frac{\rho^3(4-\rho^2)}{(1-\rho^2)^{5/2}}\sinh^{-1}\frac{\sqrt{1-\rho^2}}{\rho}\right]$$

for oblate shape fillers($\rho < 1$) with

$$P_{1212} = \frac{1}{4}(P_{1111} + P_{2222}), \qquad P_{2323} = \frac{1}{2}P_{2222}, \qquad Q_{2233} = \frac{1}{3}Q_{2222}$$

Other components can be obtained by using the symmetry condition in the 2-3 plane as well as the minor and major symmetry of $P$ and $Q$ tensors. In the limit of zero spring compliance (i.e. $\alpha = 0$), the modified Eshelby tensor($S$) becomes the original Eshelby tensor($\tilde{S}$) because $R$ becomes zero tensor. The effective moduli in the mMT scheme can be obtained by replacing the original Eshelby tensor($S$) with the modified Eshelby tensor($\tilde{S}$) (Qu, 1993),

$$L_{eff} = (c_0 L_0 + c_1 L_1 : \tilde{A}) : (c_0 I + c_1 \tilde{A} + c_1 R : L_1 : \tilde{A})^{-1} \tag{9}$$

where $\tilde{A}$ is the modified local strain concentration tensor, $\tilde{A} = [I + \tilde{S} : L_0^{-1} : (L_1 - L_0)]^{-1}$.

We plot the five independent $P$ and $Q$ components as functions of the aspect ratio $\rho$ in Fig.3. We confirm that the newly derived analytic expression of $Q_{1122}$ matches with the numerical integration obtained by Gaussian quadrature. In addition, our expression matches with the readily available analytic expressions at two limiting cases of $\rho = 1$ and $\rho = \infty$ in Qu et al. (Qu, 1993). We present the validation of all other nonzero components in Fig.3. We find that there exists a mathematical error in $Q_{1122}$ component and typographical errors in the $Q_{1111}$ and $Q_{2222}$ components in the previous study (Barai and Weng, 2011; Hashemi, 2016) and that these errors were applied to the several follow-up studies (Barai and Weng, 2011; Pan et al., 2013; Shokrieh et al., 2016; Yang et al., 2013b). The uncorrected $Q_{1122}$ diverges as the aspect ratio goes to unity and is larger than the correct value in the entire range of $\rho$. Because the $R$ tensor affects modified Eshelby tensor and effective modulus, the effective moduli based on non-corrected $R$ tensor have singularities at some values of the aspect ratio.

### 2.2 Orientation average scheme

The derived mathematical expression on the effective modulus is only applicable for the composites with completely aligned fillers. However, in the realistic composites, fillers are randomly oriented or partially aligned (Fu et al., 2000; Pötschke et al., 2005; Zaixia et al., 2006) as depicted in Fig. 4. Because the originally randomly oriented fillers in the liquid-state matrix are drawn along one axis ($X_1$) in most manufacturing processes and experiments, we limit our focus on the axisymmetric orientation distribution. Following the previous studies(Odegard et al., 2003), we define the orientation averaged Mori-Tanaka (oaMT) as Eq. (10).

$$L_{eff} = (c_0 L_0 + c_1 <L_1:\tilde{A}>):(c_0 I + c_1 <\tilde{A}> + c_1 <R:L_1:\tilde{A}>)^{-1} \tag{10}$$

where the operator $<>$ denotes the orientation average of each tensor. When the orientation distribution $\lambda(\theta)$ is function of $\theta$ only, i.e. for axis-symmetry distribution, the orientation average of an arbitrary 4$^{th}$ order tensor X can be defined as

$$<X>_{ijkl} = \frac{\int_0^{\frac{\pi}{2}} \int_{-\pi}^{\pi} X'_{ijkl}(\phi,\theta)\lambda(\theta)\sin(\theta)\,d\phi d\theta}{\int_0^{\frac{\pi}{2}} \int_{-\pi}^{\pi} \lambda(\theta)\sin\theta\,d\phi d\theta} \tag{11}$$

where $\theta$ and $\phi$ are azimuthal and polar angle with respect to the global coordinate (see Fig.5). $X'$ is the tensor transformed to global coordinate system. Following the coordinate transformation rule of 4$^{th}$ order tensor with rotation matrix $c$, $X'_{ijkl}$ can be expressed as Eq.(12).

$$X'_{ijkl} = c_{ip}c_{jq}c_{kr}c_{ls}X_{pqrs} \qquad c = \begin{bmatrix} \cos\theta & -\sin\theta & 0 \\ \sin\theta\cos\phi & \cos\theta\cos\phi & -\sin\phi \\ \sin\theta\sin\phi & \cos\theta\sin\phi & \cos\phi \end{bmatrix} \tag{12}$$

The axisymmetric orientation distribution function can be categorized into three types: 3D random, normal distribution, and aligned.

$\lambda(\theta) = 1$             : 3D random distribution

$\lambda(\theta) = \exp(-k\theta^2)$    : normal distribution                                                (13)

$\lambda(\theta) = \delta(\theta)$            : aligned distribution

As visualized in Fig.6, when $k$ in the normal distribution goes to zero or infinite, the distribution converges to random or fully aligned distributions, respectively. We note that previously used $3-1-3$ Euler angle set cannot describe the axisymmetric distribution along the $X_1$ axis due to the geometrical constraint, while several studies have adopted the $3-1-3$ Euler angle set to describe the composites with axisymmetric filler orientation distribution (Alian et al., 2015; Nguyen et al., 2013; Pan et al., 2016). When $k$ goes to zero, i.e., for a random orientation distribution, the composite must behave as an isotropic material. However, as shown in Fig.7, the longitudinal and transverse Young's modulus from the previous study do not converge to the same value in the random orientation limit. In contrast, the oaMT with our orientation average scheme predicts that both longitudinal and transverse moduli approach the modulus of the composite with randomly oriented fillers, whose analytic expression is available.

The evaluation of the orientation average in Eq. (11) is rather difficult because it involves complex transformations of $4^{th}$ order tensor to global axis. Hence, no analytic expression exists on the orientation average for general axisymmetric distribution, whereas analytic expressions were derived for the two limiting cases of the 3D random and fully aligned distributions. To facilitate the use of the oaMT, we derive algebraic expressions of the orientation averaged transversely isotropic $4^{th}$ order tensor under general axis-symmetry distribution of $\lambda(\theta)$. We express the orientation average of the transversely isotropic $4^{th}$ order tensor $<A>$ under arbitrary axis-symmetry distribution function, $\lambda(\theta)$, following Hill's notation, as

$$A = (2k_0, l_0, l'_0, n_0, 2m_0, 2p_0) \rightarrow <A> = (2k_1, l_1, l'_1, n_1, 2m_1, 2p_1). \qquad (14)$$

For the transversely isotropic $4^{th}$ order material stiffness tensor, one can write Hooke's law using the six elastic constants, as follows,

$$(\sigma_{22} + \sigma_{33}) = 2k_0(\varepsilon_{22} + \varepsilon_{33}) + 2l'_0\varepsilon_{11},$$

$$\sigma_{11} = l_0(\varepsilon_{22} + \varepsilon_{33}) + n_0\varepsilon_{11},$$

$$(\sigma_{22} - \sigma_{33}) = 2n_0(\varepsilon_{22} - \varepsilon_{33}),$$

$$\sigma_{23} = 2m_0\varepsilon_{23}, \qquad \sigma_{12} = 2p_0\varepsilon_{12}, \qquad \sigma_{13} = 2p_0\varepsilon_{13}$$

(15)

We find that the six independent components of $<A>$ can be expressed as arithmetic sum of eight constants $b_0, \ldots, b_7$ which are simple integrals including $\lambda(\theta)$.

$$k_1 = \frac{1}{4b_0}\left[k_0(b_0 + b_1 + 2b_2) + l_0(b_3 + b_4) + l'_0(b_3 + b_4) + m_0(b_0 + b_1 - 2b_2) + n_0(b_5) + p_0(4b_3)\right]$$

$$l_1 = \frac{1}{2b_0}\left[k_0(b_3 + b_4) + l_0(b_1 + b_2) + l'_0(b_5) + n_0(b_3) + m_0(b_3 - b_4) + p_0(-4b_3)\right]$$

$$l'_1 = \frac{1}{2b_0}\left[k_0(b_3 + b_4) + l(b_5) + l'_0(b_1 + b_2) + n_0(b_3) + m_0(b_3 - b_4) + p_0(-4b_3)\right]$$

(16)

$$n_1 = \frac{1}{b_0}\left[k_0(b_5) + l_0(b_3) + l'_0(b_3) + n_0(b_1) + m_0(b_5) + p_0(4b_3)\right]$$

$$m_1 = \frac{1}{8b_0}\left[k_0(b_0 + b_1 - 2b_2) + l_0(b_3 - b_4) + l'_0(b_3 - b_4) + n_0(b_5) + m_0(b_0 + b_1 + 6b_2) + p_0(4b_3 + 4b_4)\right]$$

$$p_1 = \frac{1}{16b_0}\left[k_0(b_0 - b_7) + l_0(-b_0 + b_7) + l'_0(-b_0 + b_7) + n_0(b_0 - b_7) + m_0(5b_0 - 4b_6) + p_0(8b_0 + 4b_6 + 4b_7)\right]$$

where

$$b_0 = \int_0^{\frac{\pi}{2}} \lambda(\theta)\sin\theta d\theta \qquad b_1 = \int_0^{\frac{\pi}{2}} \lambda(\theta)\cos^4\theta \sin\theta d\theta \qquad b_2 = \int_0^{\frac{\pi}{2}} \lambda(\theta)\cos^2\theta \sin\theta d\theta$$

$$b_3 = \int_0^{\frac{\pi}{2}} \lambda(\theta)\cos^2\theta \sin^3\theta d\theta \qquad b_4 = \int_0^{\frac{\pi}{2}} \lambda(\theta)\sin^3\theta d\theta \qquad b_5 = \int_0^{\frac{\pi}{2}} \lambda(\theta)\sin^5\theta d\theta$$

(17)

$$b_6 = \int_0^{\frac{\pi}{2}} \lambda(\theta)\cos 2\theta d\theta \qquad b_7 = \int_0^{\frac{\pi}{2}} \lambda(\theta)\cos 4\theta d\theta$$

When $\lambda(\theta)$ is constant (3D random distribution function), the orientation average expression reproduces the analytic results of the random orientation case (Barai and Weng, 2011; Qiu and Weng, 1990; Yang et al., 2013b). We expect that our algebraic expression of orientation average can be widely applicable to the problems described by transversely isotropic 4$^{\text{th}}$ order tensors.

## 3. Results and Discussions

**Table 1** Elastic constants used in the calculation of the effective moduli of the CNT-reinforced LaRC-CP2 polymer composite. The data are based on other research studies(Barai and Weng, 2011; Odegard et al., 2003), and the tangential compliance at interfacial damage between the polymer and CNT is set 10 nm/GPa(Barai and Weng, 2011).

| Material properties | Matrix phase | Carbon nanotube (CNT) |
|---|---|---|
| Young's modulus (E) | 0.85 GPa | |
| Poisson's ratio ($\nu$) | 0.4 | |
| Longitudinal Young's modulus ($E_{11}$) | | 1.06 TPa |
| Transverse bulk modulus ($\kappa_{23}$) | | 271 GPa |
| Transverse shear modulus ($\mu_{23}$) | | 17 GPa |
| In-plane shear modulus ($\mu_{12}$) | | 442 GPa |
| In-plane Poisson's ratio ($\nu_{12}$) | | 0.162 |

We apply our oaMT with the interfacial damage correction to the prediction of the effective moduli of CNT-reinforced LaRC-CP2 polymers. The material constants used in our calculations are listed in Table 1. Following previous studies (Barai and Weng, 2011), we only consider the tangential part in the linear spring model (see Fig.2) and set the normal compliance to zero ($\beta = 0$) because finite normal spring compliance may cause a nonphysical configuration such as the overlap of the filler and matrix volumes. To avoid such a problem, it is necessary to set the spring compliance to be a non-linear function of the displacement jump, which makes the evaluation of effective moduli highly difficult (Wang et al., 2005), which is beyond the scope of this study. The tangential spring model can describe the interfacial slip.

## 3.1 Effective modulus of composite in the presence of interfacial damage

First, we compare the effective moduli of composites with the previous $R$ tensor and the corrected $R$ tensor. We correct two simple typographical errors in the $Q_{1111}$ and $Q_{2222}$ components, but leave the mathematical error in the $Q_{1122}$ component when considering the previous $R$ tensor. Fig.8 and Fig.9 depicts the elastic moduli as functions of the filler aspect ratio for the two limiting cases of fully random and fully aligned distributions respectively. Figs.8 show the Young's modulus and the shear modulus of the composite when the fillers are randomly oriented. Under perfect interfacial bonding condition, both elastic moduli increase with the aspect ratio of fillers at a given volume fraction (Fig.8). The moduli dramatically rise when the aspect ratio reaches approximately 100 and saturate when the aspect ratio exceed 1000. The same trend can be found in the longitudinal and transverse Young's modulus when fillers are fully aligned (Fig.9). We note that, when the interfacial imperfection is taken into account, the effective moduli should become lower than the perfect bonding condition. As shown in the Fig.8 and Fig.9, the effective moduli obtained from the previous $R$ tensor are lower than those with perfect bonding in the most range. However, they have singularities near some aspect ratio values due to the presence of a singularity in the wrongly obtained $Q_{1122}$ component. The singularity in the effective moduli can only be found when the aspect ratio is less than 100, and the number of singularities change with material properties of each phases and the volume fraction of fillers. The insets in Fig.8 and Fig.9 show that effective modulus at the aspect ratio of unity, i.e., $\rho = 1$. The effective moduli based on previous $R$ tensor does not converge to the effective moduli of composite with spherical fillers, i.e., $\rho = 1$ (Qu, 1993). In comparison, the effective moduli obtained from the corrected $R$ tensor are continuous functions of the aspect ratio, are always lower than their counterparts with perfect bonding condition, and converge to the known values for spherical fillers. We suspect that these problems have not been identified because previous studies consider the composites having fillers in the range of very high aspect ratio (> 1,000) where the previous $R$ tensor does not show a singularity (Barai and Weng, 2011; Pan et al., 2013; Shokrieh et al., 2016; Yang et al., 2013b). In addition, for large aspect ratio, the difference between the results with previous and corrected $R$ tensors becomes rather negligible because the $Q_{1122}$ component becomes nearly zero in the limit of large aspect ratio (Fig.3). However, we note that the use of the corrected $R$ tensor is crucial in the small aspect ratio range smaller than 100.

### 3.2 Effective modulus of composite with partially aligned fillers

Second, we compute the effective modulus when the fillers are partially aligned with the orientation distribution of $\lambda(\theta) = \exp(-k\theta^2)$. We can study the effect of alignment by tuning the parameter $k$. The orientation averaged transversely isotropic tensor in Eq. (6) can be easily evaluated from the linear combinations of eight constants. Figs.10 show the change of longitudinal and transverse Young's modulus with $k$ and aspect ratio $\rho$ at a fixed volume fraction (2.5%). Because the moduli are calculated from the corrected $R$ tensor, they do not suffer from singularity problems. When the degree of filler alignment reduces (i.e., when $k$ decreases), the difference between two moduli diminishes and they eventually converge at $k = 0$. We find that both moduli increase with the aspect ratio. In contrast, when $k$ increases, the longitudinal modulus increases whereas the transverse modulus decreases, as expected. We also demonstrate the combined effect of volume fraction and orientation distribution in Fig.11. Both moduli increase when the volume fraction increases, and they approach the modulus of matrix in the zero volume fraction limit. We note that the contour plot depicted in Fig.11 can be easily obtained for any arbitrary axisymmetric orientation distribution of axisymmetric ellipsoidal fillers by computing the eight constants $b_0, \ldots, b_7$. which will facilitate the use of the micromechanics approach in designing and predicting the performance of the composites.

## 4. Conclusions

We computed the effective moduli of composites as functions of the aspect ratio of fillers, degree of orientation in the presence of interfacial damage. With appropriate corrections in the interfacial damage tensor and orientation average scheme, the spurious singularities in the effective composite modulus predictions are eliminated, and the longitudinal and transverse moduli converges in the limit of random orientation distribution of fillers. We demonstrated that the magnitude of moduli can be tuned by either the aspect ratio or the volume fraction, and that the directional elastic properties can be tuned by controlling the orientation distribution of fillers. To facilitate the application of our method, we derived an analytic expression for the orientation average for arbitrary axisymmetric filler orientation distribution. We believe that our study can provide a comprehensive guidance in the effort to tune the effective moduli of composite materials.

**Acknowledgments**

The authors acknowledge the support of the Basic Science Research Program through the National Research Foundation of Korea (NRF) funded by the Ministry of Science, ICT & Future Planning (2016R1C1B2011979) and the support from the Agency for Defense Development of Korea (15-201-701-008).

**Figure Set**

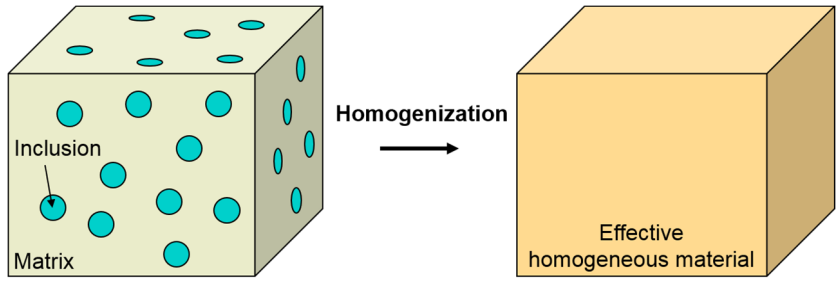

Fig.1 Schematic of homogenization scheme for RVE of reinforced composite.

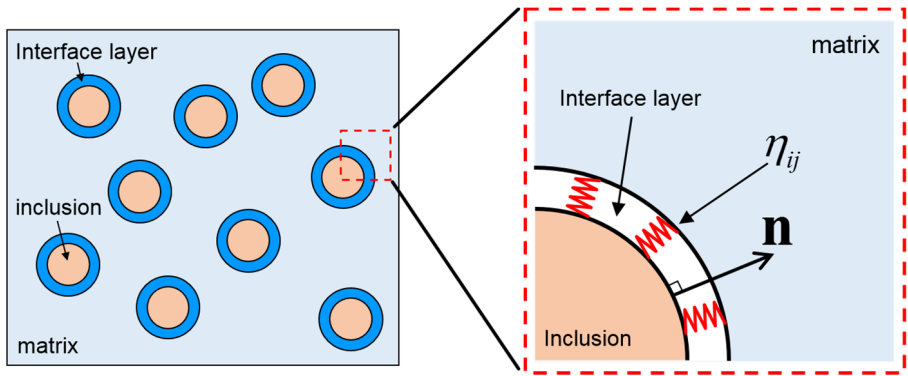

Fig.2 Interface spring model to simulate the damaged interface between matrix and filler. To visualize the interface spring, we draw a gap at the interface whose thickness is originally zero.

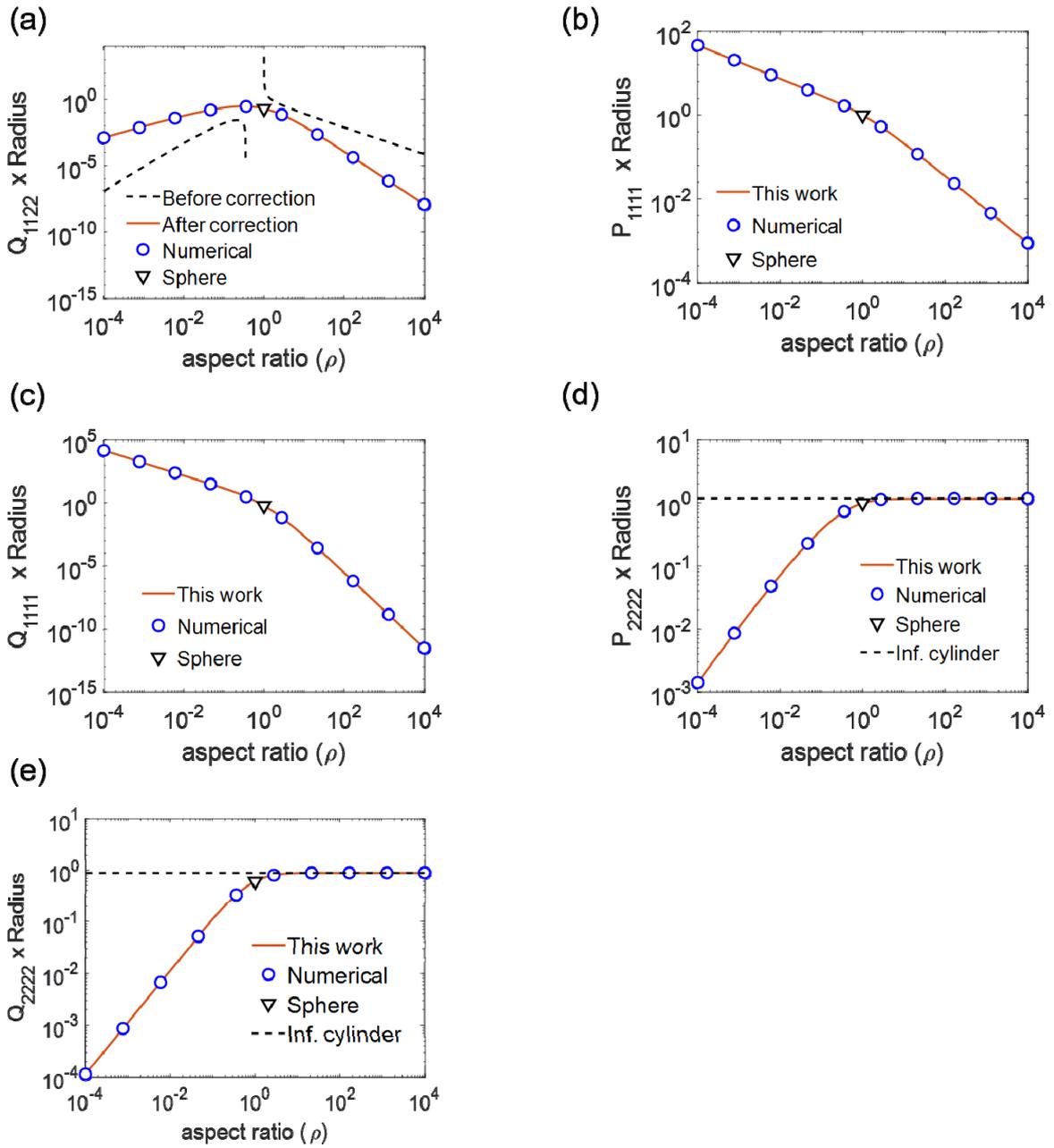

Fig.3 Independent components in *P* and *Q* tensor respect to aspect ratio of the filler. Previous and corrected $Q_{1122}$ component, one of the damaged interface properties components, with respect to the aspect ratio of the filler. The triangular symbol and dashed line are the known value at the aspect ratio of 1 and infinite respectively, reported by (Qu, 1993).

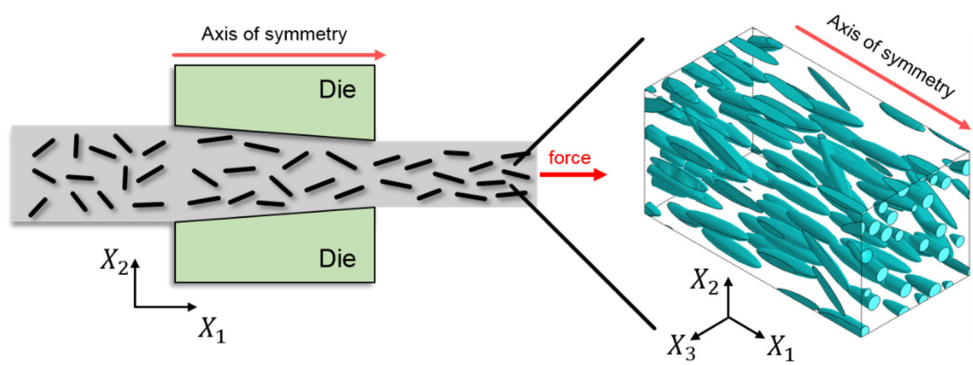

Fig.4 The drawing process at composite manufacturing process and the microstructure of composite after process.

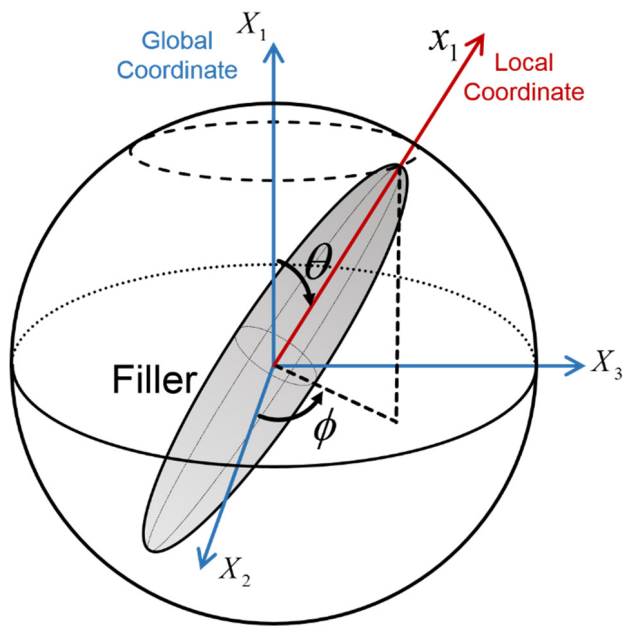

Fig.5 The local coordinates $(x_1, x_2, x_3)$ and global coordinates $(X_1, X_2, X_3)$ displayed with the spherical coordinates.

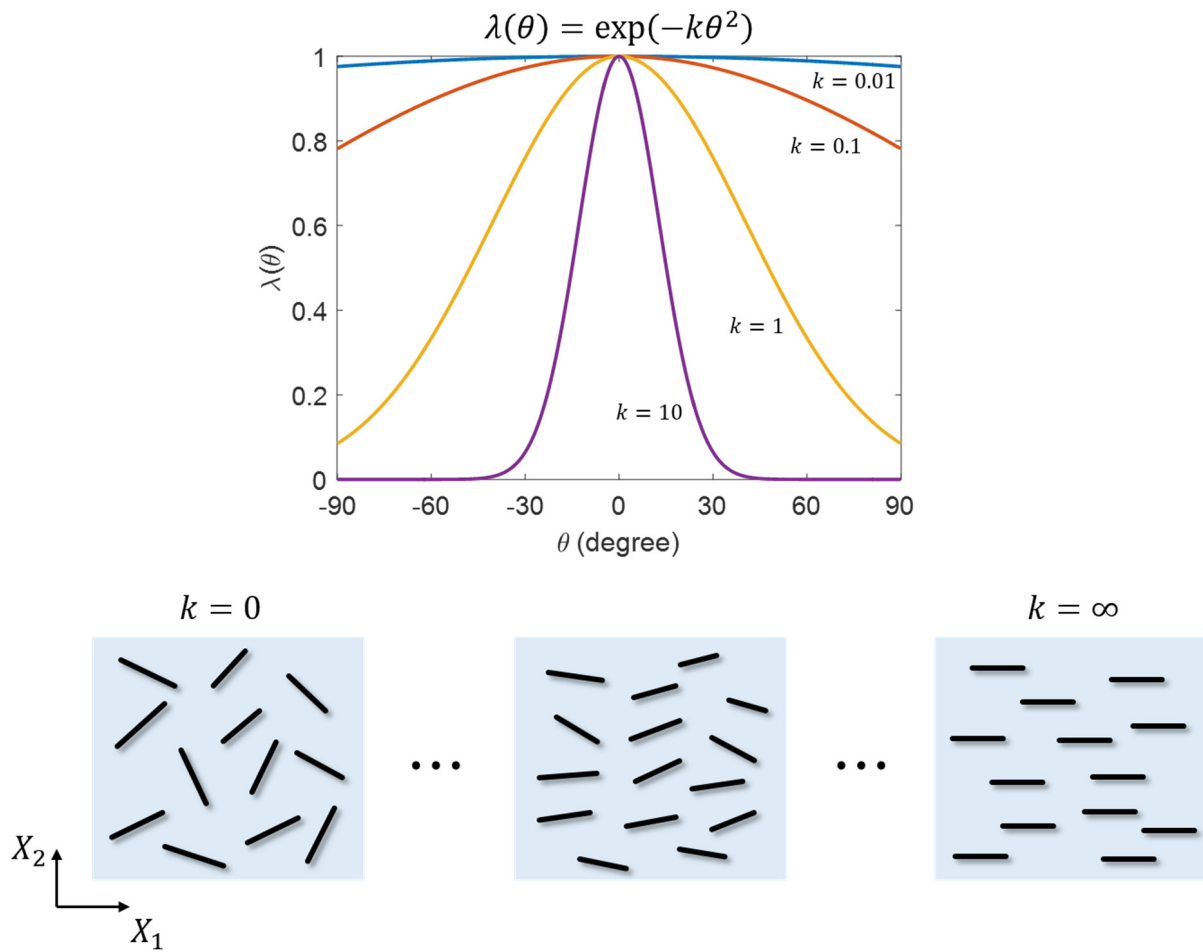

Fig.6 Various orientation distribution functions of fillers in composite respect to different *k* value and the microstructure for corresponding *k* value.

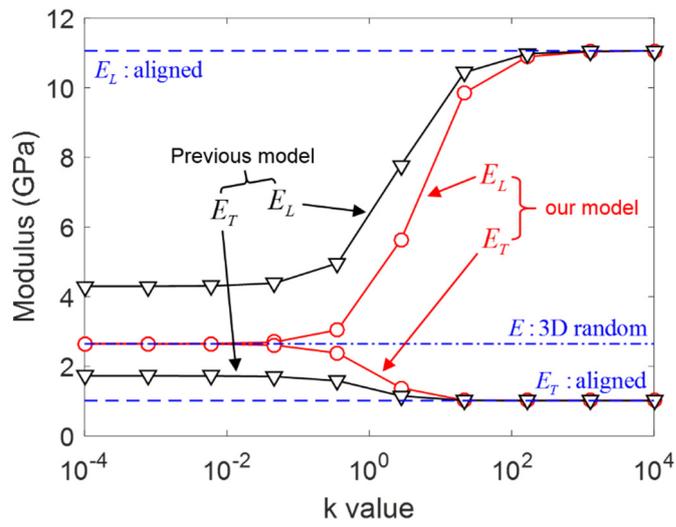

Fig.7 Modulus in the longitudinal ($X_1$) and transverse directions ($X_2, X_3$). The dashed lines are the results from known analytic expressions. The parameters used in calculation are listed in Table 1, and the volume concentration ($c_1$) is 1%.

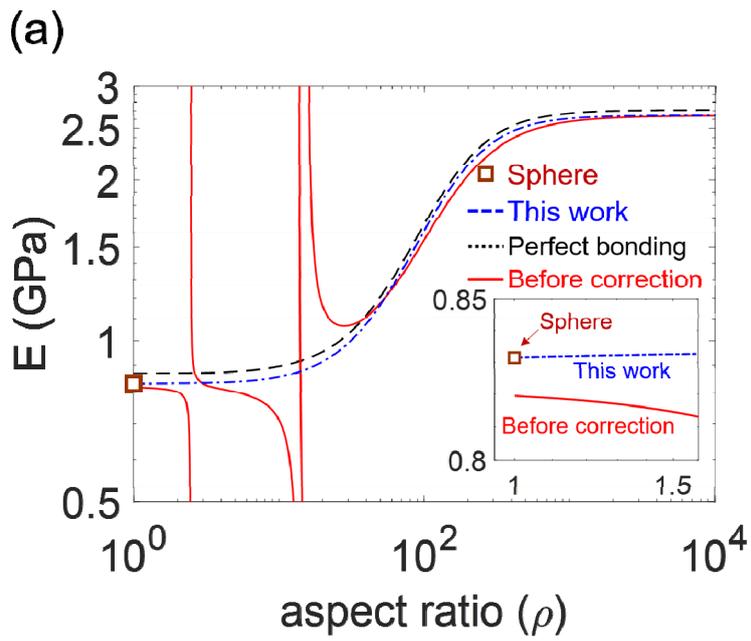

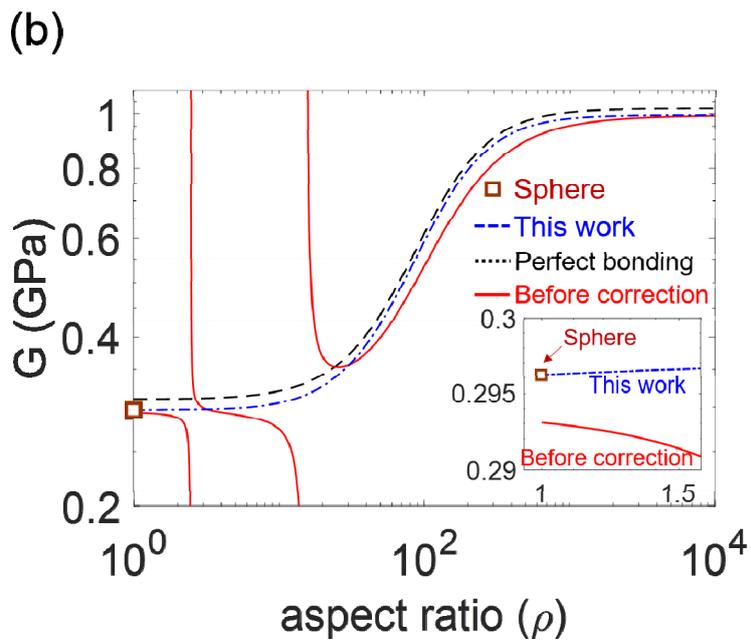

Fig.8 (a) Effective Young's modulus and (b) shear modulus of a composite with randomly oriented fillers with 1% volume concentration. The inset in each figure show the results near the aspect ratio of 1.

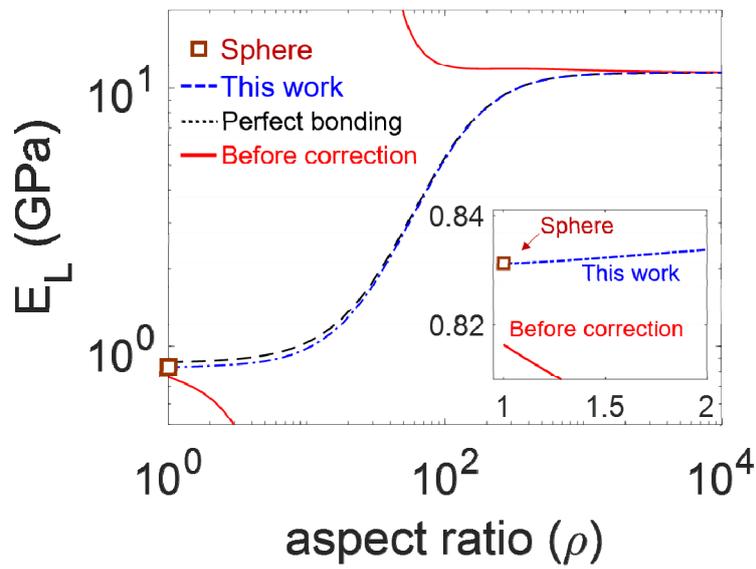

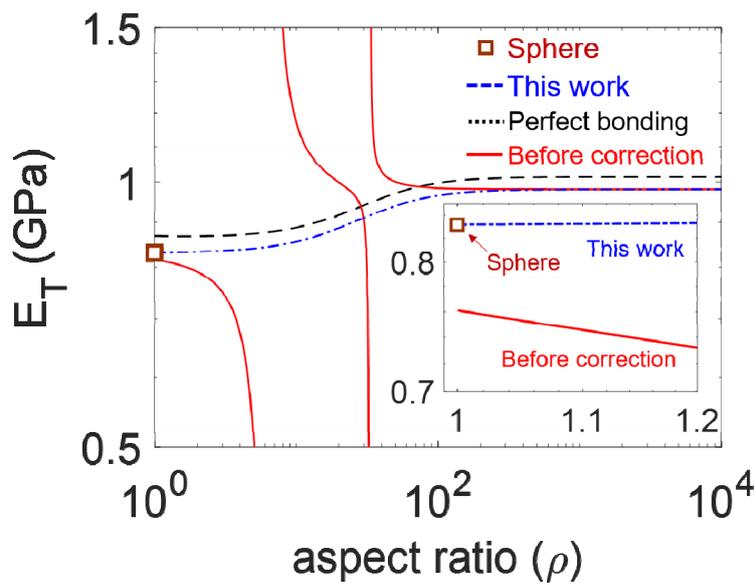

Fig.9 (a) Longitudinal Young's modulus and (b) transverse Young's modulus of a composite whose orientation distribution of fillers is aligned distribution. The volume concentration of fillers is 1%.

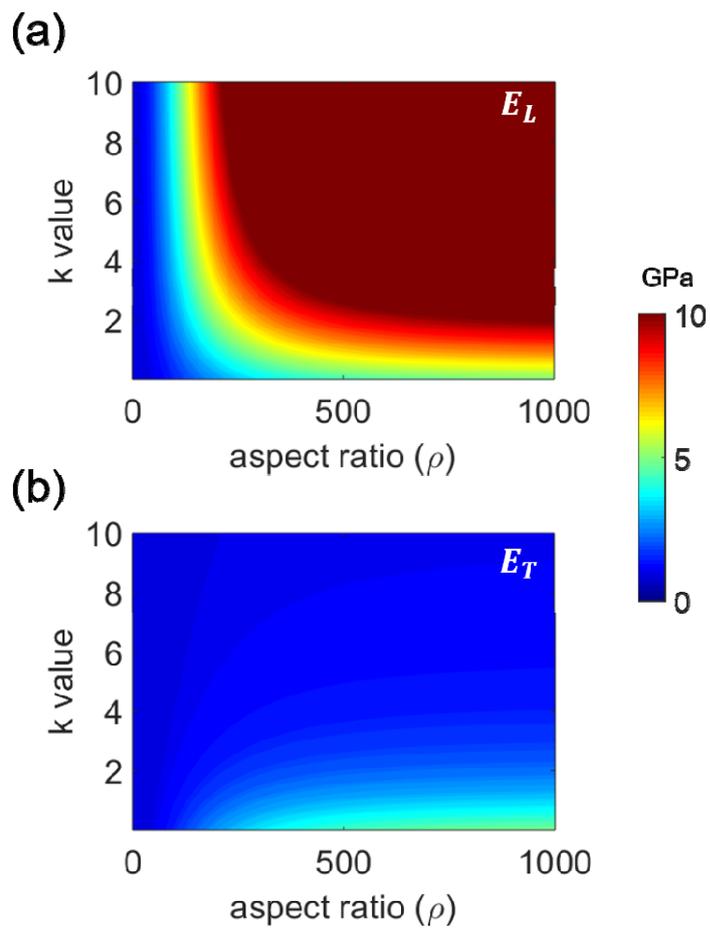

Fig.10 (a) Longitudinal Young's modulus($E_L$) and (b) transverse Young's modulus($E_T$) with a normal distribution of fillers as functions of $k$ value and aspect ratio of fillers under fixed volume fraction (2.5%)

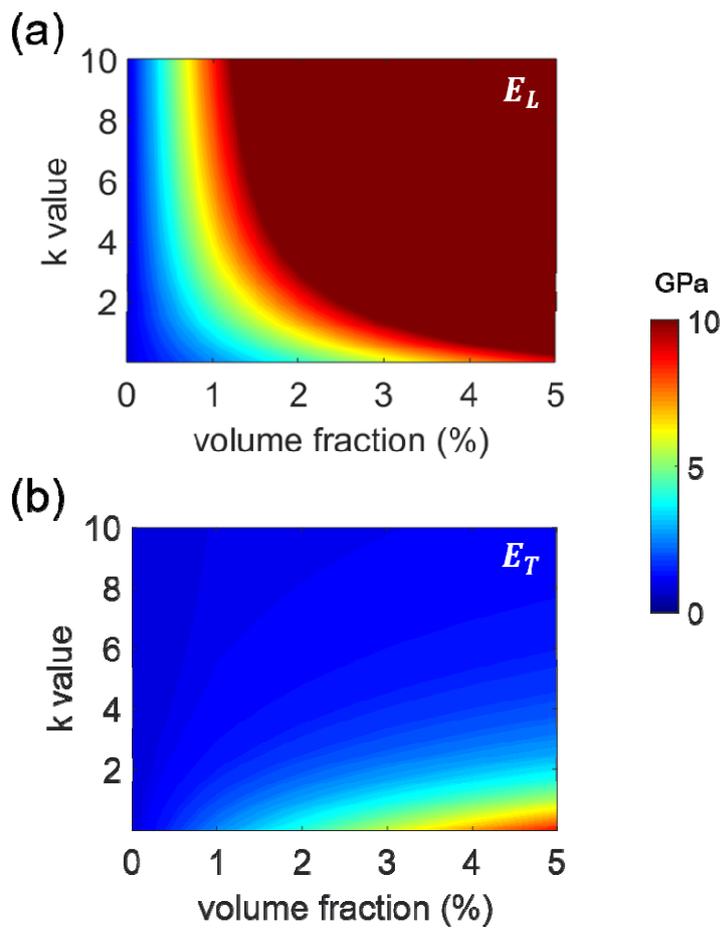

Fig.11 (a) Longitudinal ($E_L$) and (b) transverse ($E_T$) Young's modulus with respect to the degree of orientation ($k$) and the volume fraction of the fillers under fixed aspect ratio 1000.

**Appendix: Eshelby tensor (S) for ellipsoidal shape inclusion.**

For an ellipsoidal inclusion with symmetric axis(x1) in isotropic matrix, Eshelby tensor can be expressed as below,

$$S_{1111} = \frac{1}{2(1-\nu_0)}\left\{1 - 2\nu_0 + \frac{3\rho^2 - 1}{\rho^2 - 1} - \left[1 - 2\nu_0 + \frac{3\rho^2}{\rho^2 - 1}\right]g\right\}$$

$$S_{2222} = S_{3333} = \frac{3}{8(1-\nu_0)}\frac{\rho^2}{\rho^2 - 1} + \frac{1}{4(1-\nu_0)}\left[1 - 2\nu_0 - \frac{9}{4(\rho^2 - 1)}\right]g$$

$$S_{2233} = S_{3322} = \frac{1}{4(1-\nu_0)}\left\{\frac{\rho^2}{2(\rho^2 - 1)} - \left[1 - 2\nu_0 + \frac{3}{4(\rho^2 - 1)}\right]g\right\}$$

$$S_{2211} = S_{3311} = -\frac{1}{2(1-\nu_0)}\frac{\rho^2}{\rho^2 - 1} + \frac{1}{4(1-\nu_0)}\left\{\frac{3\rho^2}{\rho^2 - 1} - (1 - 2\nu_0)\right\}g$$

$$S_{1122} = S_{1133} = -\frac{1}{2(1-\nu_0)}\left[1 - 2\nu_0 + \frac{1}{\rho^2 - 1}\right] + \frac{1}{2(1-\nu_0)}\left[1 - 2\nu_0 + \frac{3}{2(\rho^2 - 1)}\right]g$$

$$S_{2323} = \frac{1}{4(1-\nu_0)}\left\{\frac{\rho^2}{2(\rho^2 - 1)} + \left[1 - 2\nu_0 - \frac{3}{4(\rho^2 - 1)}\right]g\right\}$$

$$S_{1212} = S_{1313} = \frac{1}{4(1-\nu_0)}\left\{1 - 2\nu_0 - \frac{\rho^2 + 1}{\rho^2 - 1} - \frac{1}{2}\left[1 - 2\nu_0 - \frac{3(\rho^2 + 1)}{\rho^2 - 1}\right]g\right\}$$

Where $\rho$ is the aspect ratio of filler and $\nu_0$ is Poisson ratio of matrix. Other components can be obtained by using minor symmetry condition ($S_{ijkl} = S_{jikl} = S_{ijlk}$). The $g$ is given by

$$g = \frac{\rho}{(\rho^2 - 1)^{3/2}}\left\{\rho(\rho^2 - 1)^{\frac{1}{2}} - \cosh^{-1}\rho\right\} \qquad \text{prolate shape}$$

$$= \frac{\rho}{(1 - \rho^2)^{3/2}}\left\{\cos^{-1}\rho - \rho(1 - \rho^2)^{\frac{1}{2}}\right\} \qquad \text{oblate shape}$$

# References

Alian, A.R., Kundalwal, S.I., Meguid, S.A., 2015. Interfacial and mechanical properties of epoxy nanocomposites using different multiscale modeling schemes. Composite Structures 131, 545-555.

Amjadi, M., Pichitpajongkit, A., Lee, S., Ryu, S., Park, I., 2014. Highly Stretchable and Sensitive Strain Sensor Based on Silver Nanowire–Elastomer Nanocomposite. ACS Nano 8, 5154-5163.

Banerjee, S., Sankar, B.V., 2014. Mechanical properties of hybrid composites using finite element method based micromechanics. Composites Part B: Engineering 58, 318-327.

Barai, P., Weng, G.J., 2011. A theory of plasticity for carbon nanotube reinforced composites. International Journal of Plasticity 27, 539-559.

Buck, F., Brylka, B., Müller, V., Müller, T., Weidenmann, K.A., Hrymak, A.N., Henning, F., Böhlke, T., 2015. Two-scale structural mechanical modeling of long fiber reinforced thermoplastics. Composites Science and Technology 117, 159-167.

Cui, G., Gu, L., Zhi, L., Kaskhedikar, N., van Aken, P.A., Müllen, K., Maier, J., 2008. A Germanium–Carbon Nanocomposite Material for Lithium Batteries. Advanced Materials 20, 3079-3083.

Dinzart, F., Sabar, H., 2017. New micromechanical modeling of the elastic behavior of composite materials with ellipsoidal reinforcements and imperfect interfaces. International Journal of Solids and Structures 108, 254-262.

Dunn, M.L., Ledbetter, H., Heyliger, P.R., Choi, C.S., 1996. Elastic constants of textured short-fiber composites. Journal of the Mechanics and Physics of Solids 44, 1509-&.

Eshelby, J.D., 1957. The Determination of the Elastic Field of an Ellipsoidal Inclusion, and Related Problems. Proceedings of the Royal Society of London A: Mathematical, Physical and Engineering Sciences 241, 376-396.

Fan, Z., Advani, S.G., 2005. Characterization of orientation state of carbon nanotubes in shear flow. Polymer 46, 5232-5240.

Fu, S.Y., Lauke, B., Mäder, E., Yue, C.Y., Hu, X., 2000. Tensile properties of short-glass-fiber- and short-carbon-fiber-reinforced polypropylene composites. Composites Part A: Applied Science and Manufacturing 31, 1117-1125.

Hashemi, R., 2016. On the overall viscoelastic behavior of graphene/polymer nanocomposites with imperfect interface. International Journal of Engineering Science 105, 38-55.

Hill, R., 1965. A self-consistent mechanics of composite materials. Journal of the Mechanics and Physics of Solids 13, 213-222.

Huang, H., Yin, S.C., Kerr, T., Taylor, N., Nazar, L.F., 2002. Nanostructured Composites: A High Capacity, Fast Rate Li3V2(PO4)3/Carbon Cathode for Rechargeable Lithium Batteries. Advanced Materials 14, 1525-1528.

Jiang, B., Liu, C., Zhang, C., Wang, B., Wang, Z., 2007. The effect of non-symmetric distribution of fiber orientation and aspect ratio on elastic properties of composites. Compos Part B-Eng 38, 24-34.

Kundalwal, S.I., Kumar, S., 2016. Multiscale modeling of stress transfer in continuous microscale fiber reinforced composites with nano-engineered interphase. Mechanics of Materials 102, 117-131.

Lee, S., Amjadi, M., Pugno, N., Park, I., Ryu, S., 2015. Computational analysis of metallic nanowire-elastomer nanocomposite based strain sensors. AIP Advances 5, 117233.

Lü, C., Cheng, Y., Liu, Y., Liu, F., Yang, B., 2006. A Facile Route to ZnS–Polymer Nanocomposite Optical Materials with High Nanophase Content via γ-Ray Irradiation Initiated Bulk Polymerization. Advanced Materials 18, 1188-1192.

Lü, C., Guan, C., Liu, Y., Cheng, Y., Yang, B., 2005. PbS/Polymer Nanocomposite Optical Materials with High Refractive Index. Chemistry of Materials 17, 2448-2454.

Luding, S., 2008. Cohesive, frictional powders: contact models for tension. Granular Matter 10, 235.

Nguyen, H.G., Ortola, S., Ghorbel, E., 2013. Micromechanical modelling of the elastic behaviour of polymer mortars. European Journal of Environmental and Civil Engineering 17, 65-83.

Nyholm, L., Nyström, G., Mihranyan, A., Strømme, M., 2011. Toward Flexible Polymer and Paper-Based Energy Storage Devices. Advanced Materials 23, 3751-3769.

Odegard, G.M., Gates, T.S., Wise, K.E., Park, C., Siochi, E.J., 2003. Constitutive modeling of nanotube–reinforced polymer composites. Composites Science and Technology 63, 1671-1687.

Oh, S.-M., Oh, S.-W., Yoon, C.-S., Scrosati, B., Amine, K., Sun, Y.-K., 2010. High-Performance Carbon-LiMnPO4 Nanocomposite Cathode for Lithium Batteries. Advanced Functional Materials 20, 3260-3265.

Pan, J., Bian, L., Zhao, H., Zhao, Y., 2016. A new micromechanics model and effective elastic modulus of nanotube reinforced composites. Computational Materials Science 113, 21-26.

Pan, Y., Weng, G.J., Meguid, S.A., Bao, W.S., Zhu, Z.H., Hamouda, A.M.S., 2013. Interface effects on the viscoelastic characteristics of carbon nanotube polymer matrix composites. Mechanics of Materials 58, 1-11.

Pötschke, P., Brünig, H., Janke, A., Fischer, D., Jehnichen, D., 2005. Orientation of multiwalled carbon nanotubes in composites with polycarbonate by melt spinning. Polymer 46, 10355-10363.

Pushparaj, V.L., Shaijumon, M.M., Kumar, A., Murugesan, S., Ci, L., Vajtai, R., Linhardt, R.J., Nalamasu, O., Ajayan, P.M., 2007. Flexible energy storage devices based on nanocomposite paper. Proceedings of the National Academy of Sciences 104, 13574-13577.

Qiu, Y.P., Weng, G.J., 1990. On the application of Mori-Tanaka's theory involving transversely isotropic spheroidal inclusions. International Journal of Engineering Science 28, 1121-1137.

Qu, J., 1993. The effect of slightly weakened interfaces on the overall elastic properties of composite materials. Mechanics of Materials 14, 269-281.

Qu, J., Cherkaoui, M., 2006. Fundamentals of Micromechanics of Solids. John Wiley & Sons, Inc.

Shokrieh, M.M., Ghajar, R., Shajari, A.R., 2016. The effect of time-dependent slightly weakened interface on the viscoelastic properties of CNT/polymer nanocomposites. Composite Structures 146, 122-131.

Sodhani, D., Reese, S., 2014. Finite Element-Based Micromechanical Modeling of Microstructure Morphology in Filler-Reinforced Elastomer. Macromolecules 47, 3161-3169.

Sun, H., Di, S., Zhang, N., Wu, C., 2001. Micromechanics of composite materials using multivariable finite element method and homogenization theory. International Journal of Solids and Structures 38, 3007-3020.

Wang, J., Duan, H.L., Zhang, Z., Huang, Z.P., 2005. An anti-interpenetration model and connections between interphase and interface models in particle-reinforced composites. International Journal of Mechanical Sciences 47, 701-718.

Withers, P.J., Stobbs, W.M., Pedersen, O.B., 1989. The application of the eshelby method of internal stress determination to short fibre metal matrix composites. Acta Metallurgica 37, 3061-3084.

Yang, B.J., Shin, H., Lee, H.K., Kim, H., 2013a. A combined molecular dynamics/micromechanics/finite element approach for multiscale constitutive modeling of nanocomposites with interface effects. Applied Physics Letters 103, 241903.

Yang, S., Yu, S., Ryu, J., Cho, J.-M., Kyoung, W., Han, D.-S., Cho, M., 2013b. Nonlinear multiscale modeling approach to characterize elastoplastic behavior of CNT/polymer nanocomposites considering the interphase and interfacial imperfection. International Journal of Plasticity 41, 124-146.

Zaixia, F., Zhangyu, Yanmo, C., Hairu, L., 2006. Effects of Pre-stretching on the Tensile Properties of Knitted Glass Fiber Fabric Reinforced Polypropylene Composite. Journal of Thermoplastic Composite Materials 19, 399-411.